\documentclass[12pt]{article}
\usepackage{axodraw}
\usepackage{amssymb}
\usepackage{cite}
\usepackage[]{hyperref}
\input xy
\xyoption{all}

\begin{document}

\vspace*{0.5in}

\begin{center}

{\large\bf Duality group actions on fermions}

\vspace{0.25in}
                                                                               
Tony Pantev$^1$ and Eric Sharpe$^{2}$ \\

\vspace*{0.2in}

\begin{tabular}{cc}
{ \begin{tabular}{l}
$^1$ Department of Mathematics \\
University of Pennsylvania \\
David Rittenhouse Lab.\\
209 South 33rd Street \\
Philadelphia, PA  19104-6395
\end{tabular} } &
{\begin{tabular}{l}
$^2$ Department of Physics MC 0435\\
850 West Campus Drive\\
Virginia Tech \\
Blacksburg, VA  24061 
\end{tabular} } 
\end{tabular}

{\tt tpantev@math.upenn.edu},
{\tt ersharpe@vt.edu} \\

$\,$

\end{center}

In this short paper we look at the action of T-duality and string duality
groups on fermions, in maximally-supersymmetric theories and related
theories.
Briefly, we argue that typical duality groups such as $SL(2,{\mathbb Z})$
have sign ambiguities in their actions on fermions, 
and propose that pertinent
duality groups be extended by ${\mathbb Z}_2$, to groups such as the
metaplectic group.
Specifically, we look at duality groups arising from mapping class groups
of tori in M theory compactifications, T-duality, ten-dimensional type IIB
S-duality, and (briefly) four-dimensional $N=4$ super Yang-Mills, 
and in each case, propose that the full duality group is a nontrivial
${\mathbb Z}_2$
extension of the duality group acting on bosonic degrees of freedom,
to more accurately
describe possible actions on fermions.  We also walk through
U-duality groups for toroidal compactifications to nine, eight, and seven
dimensions, which enables us to perform cross-consistency tests of 
these proposals.

\begin{flushleft}
August 2016
\end{flushleft}

\newpage

\tableofcontents

\newpage

\section{Introduction}

In this paper we re-examine duality groups in high-dimensional string theories
with maximal supersymmetry, taking a close look at duality group 
actions\footnote{
To be clear, although duality transformations should define a map
of gauge-invariant
local operators, their action on fundamental fields
might not be so simply defined.  For one example, in four-dimensional
$N=1$ Seiberg duality, mesons are composite operators on one side and
fundamental fields on the other.
However, in maximally-supersymmetric
theories, actions on fields in low-energy effective
actions are typically well-defined, and it is for this
reason that one can {\it e.g.} describe the R-R and NS-NS two-form fields
in ten-dimensional IIB supergravity as transforming in a doublet of the
S-duality group.  For simplicity, in this note we will largely focus
on maximally-supersymmetric theories and related cases.
}
 on
fermions in low-energy effective field theories.
Historically, such duality group actions have been
primarily discussed only on bosonic degrees of freedom.
Briefly, we argue that in many cases, fermions are not uniquely defined
under the duality groups as they are typically described, because of
square-root-type branch cut ambiguities, and so propose that
those duality groups 
be slightly enlarged.  

As one prototypical example, in ten-dimensional type IIB string theory,
we argue that under the S-duality group $SL(2,{\mathbb Z})$, the
transformation of the fermions is not uniquely defined due to a sign
ambiguity, and so propose that
$SL(2,{\mathbb Z})$ should be replaced by a ${\mathbb Z}_2$ extension.
In particular, from the form of the duality group action, we propose that
$SL(2,{\mathbb Z})$ should be replaced by a particular 
nontrivial ${\mathbb Z}_2$
central extension known as the metaplectic group and denoted
$Mp(2,{\mathbb Z})$.

We examine several different duality groups -- mapping class groups of tori
arising in M theory compactifications,
T-dualities, ten-dimensional IIB S-duality -- proposing such extensions
in each case, as well as corresponding U-duality groups, which we use
to 
perform cross-consistency
tests of these proposed extensions.

We begin in section~\ref{sect:review} by briefly reviewing the metaplectic
group
$Mp(2,{\mathbb Z})$, which will be the most commonly appearing proposed
duality group.  We review its relation to $SL(2,{\mathbb Z})$ and properties of
this and related groups.  

In section~\ref{sect:mappingclass-tori}, we go
on to discuss the linear diffeomorphism mapping class groups of tori,
that play a crucial role in duality symmetries of toroidal compactifications
of M theory.  The mapping class group for an $n$-torus is ordinarily given as
$SL(n,{\mathbb Z})$, but we find that this group has an ambiguous action
on fermions, and so we propose that in general it be replaced by a 
${\mathbb Z}_2$ extension which we denote $\widetilde{SL}(n,{\mathbb Z})$.

In section~\ref{sect:tduality} we turn to perturbative
T-duality groups of toroidally-compactified string theories.
There, we argue that the worldsheet fermions themselves are well-defined
under target-space T-dualities; the Ramond sector vacua, however,
are only well-defined under ${\mathbb Z}_2$ extensions of the ordinary
duality groups, just as in the discussion of mapping class groups of
tori.  We discuss moduli spaces of SCFTs, and argue for similar
reasons that, for example, the moduli space of elliptic curves SCFTs
is best described as a quotient of the upper half plane by
$Mp(2,{\mathbb Z})$ instead of $SL(2,{\mathbb Z})$ or $PSL(2,{\mathbb Z})$.

In section~\ref{sect:10d-sduality}, 
we turn to S-duality in ten-dimensional type IIB
string theory.  Here, no tori are manifest; nevertheless, we shall argue
that the ten-dimensional fermions are also slightly ambiguous under
the action of the S-duality group $SL(2,{\mathbb Z})$, due to square-root
branch cut ambiguities, and so for a well-defined action, we propose to
promote the
S-duality group to the ${\mathbb Z}_2$-extension $Mp(2,{\mathbb Z})$.

In section~\ref{sect:4dN=4}, we briefly turn to four-dimensional
$N=4$ super Yang-Mills, and discuss how the $Mp(2,{\mathbb Z})$
action in ten-dimensional type IIB appears to imply an analogous extension
in S-duality in the four-dimensional $N=4$ theory.  That said, for most of
this paper, in order to be able to speak meaningfully about duality group
actions on fields rather than theories, we focus on maximally-supersymmetric
theories in high dimensions.

Finally, in section~\ref{sect:uduality}, we put the ideas of the proceeding
sections together to consider U-duality groups of toroidally-compactified
M theory in nine, eight, and seven dimensions.  We see that the various
occurrences of the metaplectic group and analogous ${\mathbb Z}_2$
extensions check one another.  For example, in nine dimensions, in order
for the U-duality group to be consistent, both the mapping class group
$SL(2,{\mathbb Z})$ of $T^2$ as well as the S-duality group of ten-dimensional
type IIB must be extended to $Mp(2,{\mathbb Z})$, which is what we have
proposed in previous sections.

One of the motivations for this work is to understand the physical
significance of a result in \cite{mebw}.  Specifically, it was argued
there that the moduli stack of elliptic curves for which the Bagger-Witten
line bundle is well-defined is the stacky quotient
\begin{displaymath}
\left[ \left( \mbox{upper half plane}\right) / Mp(2,{\mathbb Z}) \right],
\end{displaymath}
for $Mp(2,{\mathbb Z})$ the metaplectic group extending $SL(2,{\mathbb Z})$
by ${\mathbb Z}_2$, rather than a quotient by $SL(2,{\mathbb Z})$ or
$PSL(2,{\mathbb Z})$.  One could naturally ask whether the appearance of
the metaplectic group was merely an obscure mathematical
quirk of the Bagger-Witten line
bundle in that circumstance, or reflected the true T-dualities
of the theory.  We propose in this paper that the latter is the case.

The proposal of section~\ref{sect:10d-sduality}, that the fermions of
ten-dimensional IIB supergravity transform under $Mp(2,{\mathbb Z})$,
has been independently obtained by D.~Morrison \cite{davepriv}.

\section{Review of the metaplectic group}
\label{sect:review}

We will frequently encounter the metaplectic group $Mp(2,{\mathbb Z})$ and
its various cousins in this paper, so let us briefly review some pertinent
facts.  The metaplectic group $Mp(2,{\mathbb Z})$ is the unique nontrivial
central extensive of ${\mathbb Z}_2$, and can be described as the group
with elements of the form
\begin{displaymath}
\left( 
\left[ \begin{array}{cc}
a & b \\
c & d \end{array} \right], \:
\pm \sqrt{c \tau + d } \right),
\end{displaymath}
where
\begin{displaymath}
\left[ \begin{array}{cc}
a & b \\
c & d \end{array} \right] \: \in \: SL(2,{\mathbb Z}),
\end{displaymath}
and $\sqrt{c \tau + d}$ is considered as a holomorphic function of
$\tau$ in the upper half plane.  The multiplication is defined as
\begin{displaymath}
(A, f(\cdot)) (B, g(\cdot)) \: = \: ( AB, f(B(\cdot)) g(\cdot) ) .
\end{displaymath}

More generally, there is a metaplectic group $Mp(2k,{\mathbb Z})$ which 
is the unique nontrivial ${\mathbb Z}_2$ central extension of
the symplectic group
$Sp(2k,{\mathbb Z})$.  For $k=1$, $Sp(2k,{\mathbb Z}) = SL(2,{\mathbb Z})$,
and so this description reduces to the one above.

Metaplectic groups over ${\mathbb R}$ also define the symplectic analogue
of spin structures on oriented Riemannian manifolds (see {\it e.g.}
\cite{forgerhess} and references therein, of which we shall give
only a very brief account here).  There is a formal
definition of a metaplectic structure on a symplectic manifold,
which is an equivariant lift of the symplectic frame bundle.
Just as in ordinary spin structures, a metaplectic structure exists on a
symplectic manifold $(X,\omega)$ if and only if the second Stiefel-Whitney
class of $M$ vanishes.  Although we will not use metaplectic structures
{\it per se} in this paper, we will often see metaplectic groups and
related extensions arise via a need to define spinors in given
situations.

\section{Mapping class groups of tori}
\label{sect:mappingclass-tori}

In this section, we will argue that under the action of the
`mapping class group' $SL(n,{\mathbb Z})$ of
a torus, describing the (analogues of) Dehn twists,
spinors transform under the action of a ${\mathbb Z}_2$
extension of $SL(n,{\mathbb Z})$ (and the spin structures are 
permuted).  We will argue this solely from the torus itself;
in section~\ref{sect:tduality-worldsheet}, 
we will describe how the same result appears in the
worldsheet theory of a sigma model on a torus. 

To be clear, let us define more precisely what we mean by the `mapping
class group' of a torus.  We follow the language and conventions of
\cite{edduality}[section 3.4].  If we describe $T^n$ as a set of
real variables $y^i$ subject to $y^i \equiv y^i + n^i$ for $n^i \in
{\mathbb Z}$, then the `mapping class group' to which we refer is
the group of linear orientation-preserving diffeomorphisms
$y^i \: \mapsto \: w^i_j y^j$.  Clearly the $(w^i_j) \in GL(n,{\mathbb Z})$,
and to preserve a choice of orientation, we must require
$(w^i_j) \in SL(n,{\mathbb Z})$.

In this language, we can now begin to gain some intuition for
the subtlety that arises when
considering fermions.  Broadly speaking, the mapping class group
is acting by rotations.  However, to define an action on fermions,
one must lift to a Spin cover.  As a result, one should expect
intuitively that the mapping class group will have to be replaced\footnote{
We would like to thank D.~Auroux for a discussion of this matter.
} by
some sort of ${\mathbb Z}_2$ cover.  In this section, that is exactly
the conclusion we shall reach.  For $n=2$, the ${\mathbb Z}_2$
cover of $SL(2,{\mathbb Z})$ will be the metaplectic group
$Mp(2,{\mathbb Z})$, and for $n \geq 3$, it will be a ${\mathbb Z}_2$
cover of $SL(n,{\mathbb Z})$ which we will denote 
$\widetilde{SL}(n,{\mathbb Z})$.

\subsection{Elliptic curves}

Let us begin by considering elliptic curves.  Under the action of
the group of Dehn twists
$SL(2,{\mathbb Z})$, if we describe the elliptic curve as 
\begin{displaymath}
H / ( {\mathbb Z} + {\mathbb Z} \tau ),
\end{displaymath}
for $H$ the upper half plane, then at the same time that
\begin{displaymath}
\tau \: \mapsto \: \frac{a \tau + b}{c \tau + d},
\end{displaymath}
for
\begin{displaymath}
\left[ \begin{array}{cc}
a & b \\
c & d \end{array} \right] \: \in \: SL(2,{\mathbb Z}),
\end{displaymath}
a point $z$ in the plane transforms
as\footnote{
Briefly, the idea is that if we construct a family of elliptic curves with
complex structure parameter $\tau$ as $( {\mathbb C} \times H)/{\mathbb Z}^2$,
for $H$ the upper half plane and ${\mathbb Z}^2$ action given by
\begin{displaymath}
(m,n): \: (z, \tau) \: \mapsto \: (z + m \tau + n, \tau),
\end{displaymath}
then to be consistent with the ${\mathbb Z}^2$ action,
under $\tau \mapsto (a \tau + b)/(c \tau + d)$, we must rescale $z$, so that
\begin{displaymath}
(z, \tau) \: \cong \: \left( \frac{z}{c \tau + d}, \frac{a \tau + b}{
c \tau + d} \right),
\end{displaymath}
making the right-hand side well-defined under the
${\mathbb Z}^2$ quotient.
As consistency tests, note that under the map on $z$,
\begin{eqnarray*}
\tau & \mapsto & \frac{\tau}{c \tau + d} \: = \: \frac{a \tau + b}{c \tau + d}
+ (d-1)\left( \frac{a \tau + b}{c \tau + d} \right) - b \: \cong \:
 \frac{a \tau + b}{c \tau + d},
\\
1 & \mapsto & \frac{1}{c \tau + d} \: = \: 1 -c \left( \frac{a \tau+b}{
c \tau + d} \right) + (a-1) \: \cong \: 1.
\end{eqnarray*}
} \cite{hainrev}[section 2.3]
\begin{displaymath}
z \: \mapsto \: (c \tau + d)^{-1} \, z,
\end{displaymath}
and so the holomorphic top-form (a one-form) transforms as
\begin{displaymath}
dz \: \mapsto \: (c \tau + d )^{-1} \, dz.
\end{displaymath}
In passing, note that although the parameter $\tau$ is invariant
under the center $\{\pm 1\} \subset SL(2,{\mathbb Z})$,
the holomorphic top-form $dz$ is not invariant under the center, and so is
only uniquely defined on a (stacky) $SL(2,{\mathbb Z})$ quotient of the upper
half plane, not a 
$PSL(2,{\mathbb Z}) = SL(2,{\mathbb Z})/\{\pm 1\}$ quotient.

Now, a chiral spinor on an elliptic curve is a section of a square root
of the canonical bundle, and so should transform in the same fashion as
$\sqrt{dz}$.  
Thus, if $\psi$ is a chiral spinor, then under $SL(2,{\mathbb Z})$,
\begin{displaymath}
\psi \: \mapsto \: \pm \sqrt{ c \tau + d }^{-1} \, \psi.
\end{displaymath}
However, the element of $SL(2,{\mathbb Z})$ does not uniquely determine the
sign:  the group that is really acting is not $SL(2,{\mathbb Z})$, but
a ${\mathbb Z}_2$ extension of $SL(2,{\mathbb Z})$, and in fact
it should be clear to the reader that the ${\mathbb Z}_2$ extension in
question is $Mp(2,{\mathbb Z})$.

\subsection{Higher-dimensional tori}

So far we have argued that on a single $T^2$, the action of the mapping
class group $SL(2,{\mathbb Z})$ on fermions is ambiguous up to a 
$\tau$-dependent phase, and so the action on fermions is more properly
described as an action of $Mp(2,{\mathbb Z})$, the unique nontrivial
central extension of $SL(2,{\mathbb Z})$.

Now, consider a higher-dimensional torus $T^n$.  Setting aside fermions,
the mapping class group of such a torus is $SL(n,{\mathbb Z})$.
Now, For any $T^2 \subset
T^n$, the same considerations must apply to fermions, and so every
$SL(2,{\mathbb Z}) \subset SL(n,{\mathbb Z})$ must be enhanced to
$Mp(2,{\mathbb Z})$.

Therefore, to describe the action of the mapping class group on fermions,
we need a ${\mathbb Z}_2$ extension of $SL(n,{\mathbb Z})$, such that
every $SL(2,{\mathbb Z})$ is extended to $Mp(2,{\mathbb Z})$.

We will construct this extension, which we shall denote
$\widetilde{SL}(n,{\mathbb Z})$, for $n \geq 3$ as a pullback of the
universal cover $\widetilde{SL}(n,{\mathbb R})$ of
$SL(n,{\mathbb R})$.  

First, let us consider the universal cover $\widetilde{SL}(n,{\mathbb R})$.
Since the maximal compact subgroup of $SL(n,{\mathbb R})$ is
$SO(n)$, 
\begin{displaymath}
\pi_1( SL(n,{\mathbb R})) \: = \: \pi_1(SO(n)) \: = \: 
\left\{ \begin{array}{cl}
{\mathbb Z}_2 & n \geq 3, \\
{\mathbb Z} & n=2,
\end{array} \right.
\end{displaymath}
thus the universal cover $\widetilde{SL}(n,{\mathbb R})$ is a 
${\mathbb Z}_2$ central extension of $SL(n,{\mathbb R})$ for
$n \geq 3$, and a ${\mathbb Z}$-fold central extension for $n=2$.
We can understand this topologically as follows.  As a topological
space, $SL(n,{\mathbb R})$ is homeomorphic to $SO(n) \times {\mathbb R}^k$
for some $k$ (ignoring the group structure), so its universal cover
is homeomorphic to ${\rm Spin}(n) \times {\mathbb R}^k$.
For $n=2$, Spin$(2) = {\mathbb R}$, a ${\mathbb Z}$-fold cover,
and for $n \geq 3$, Spin$(n)$ is a double cover of $SO(n)$.

In any event, we can now construct our desired group, that extends the
action of the mapping class group to fermions.
Define $\widetilde{SL}(n,{\mathbb Z})$ by the pullback square
\begin{displaymath}
\xymatrix{
\widetilde{SL}(n,{\mathbb R}) \ar[r]^p & SL(n,{\mathbb R}) \\
\widetilde{SL}(n,{\mathbb Z}) \ar[r] \ar[u] &
SL(n,{\mathbb Z}),  \ar[u]_i
}
\end{displaymath}
where $p: \widetilde{SL}(n,{\mathbb R}) \rightarrow SL(n,{\mathbb R})$
is the projection map, and $i: SL(n,{\mathbb Z}) \hookrightarrow
SL(n,{\mathbb R})$ is inclusion.
In other words, we define 
\begin{displaymath}
\widetilde{SL}(n,{\mathbb Z}) \: = \: \left\{ 
(a,b) \in \widetilde{SL}(n,{\mathbb R})
\times SL(n,{\mathbb Z}) \, | \, p(a) = i(b) \right\},
\end{displaymath}
with product structure
\begin{displaymath}
(a,b) \cdot (a',b') \: \equiv \: (a a', b b'),
\end{displaymath}
which is well-defined because both $p$ and $i$ are homomorphisms, hence
\begin{displaymath}
p(a a') \: = \: p(a) p(a') \: = \: i(b) i(b') \: = \: i(b b').
\end{displaymath}

For $n \geq 3$, we claim that the group $\widetilde{SL}(n,{\mathbb Z})$
encodes the action of the mapping class group on the fermions.
For $n=2$, the desired group is the metaplectic group $Mp(2,{\mathbb Z})$,
which is quotient of $\widetilde{SL}(2,{\mathbb Z})$ by $2 {\mathbb Z}$,
a subgroup of the central ${\mathbb Z} \subset \widetilde{SL}(2,{\mathbb Z})$.

It remains to check whether the restriction of $\widetilde{SL}(n,{\mathbb Z})$
to $p^{-1}( SL(2,{\mathbb Z}) )$ for any $SL(2,{\mathbb Z}) \subset
SL(n,{\mathbb Z})$ is a trivial or nontrivial central extension.
(Since $Mp(2,{\mathbb Z})$ is the unique nontrivial central extension of
$SL(2,{\mathbb Z})$ by ${\mathbb Z}_2$, showing that 
$p^{-1}( SL(2,{\mathbb Z}))$ is a nontrivial central extension for every
$SL(2,{\mathbb Z})$ would imply 
$p^{-1}( SL(2,{\mathbb Z}) ) = Mp(2,{\mathbb Z})$
for every $SL(2,{\mathbb Z})$.)
Now, every
copy of $SL(2,{\mathbb Z})$ inside $SL(n,{\mathbb Z})$ 
comes from taking the integral points of a copy of $SL(2,{\mathbb R})$
inside $SL(n,{\mathbb R})$, so the question reduces to whether
for any copy of $SO(2)$ inside $SO(n)$, the Spin cover of $SO(n)$ splits
when restricted to that $SO(2)$.  

It can be shown that the Spin cover of $SO(n)$ does not split when restricted
to any $SO(2)$.  This follows\footnote{
One of the authors (E.S.)
would like to thank J.~Francis for outlining this argument.}
from the fact that the group homomorphism $SO(2) \rightarrow SO(n)$ 
is surjective
on the fundamental group, and the correspondence between covering spaces
and actions of the fundamental group.

As a result, the restriction of $\widetilde{SL}(n,{\mathbb Z})$ to
$p^{-1}(SL(2,{\mathbb Z}))$ for any $SL(2,{\mathbb Z}) \subset
SL(n,{\mathbb Z})$ is a nontrivial central extension, hence
\begin{displaymath}
p^{-1}( SL(2,{\mathbb Z})) \: = \: Mp(2,{\mathbb Z}),
\end{displaymath}
and so $\widetilde{SL}(n,{\mathbb Z})$ for $n \geq 3$ is the desired
group describing the mapping class group action on fermions.

\section{T-duality}
\label{sect:tduality}

\subsection{Action on worldsheet theories}
\label{sect:tduality-worldsheet}

In this subsection, we will consider the action on worldsheet fermions
from two different sources:
the action on the worldsheet
theory of T-duality on a target elliptic curve, and the action of Dehn twists
when the worldsheet itself is an elliptic curve.

Let us first consider a (2,2) supersymmetric
sigma model with target space $T^2$.  
Now, as is well-known, the set of T-dualities is larger than merely
a ${\mathbb Z}_2$ for each circle in $T^2$, as for example, those T-dualities
can be combined with Dehn twists to form a larger symmetry group.
Across both complex and K\"ahler structures on $T^2$, the T-duality
group acting on the CFT is well-known to be $O(2,2;{\mathbb Z})$.

Now, the part of the T-duality group $O(2,2;{\mathbb Z})$
that preserves GSO projections is
\begin{displaymath}
SO(2,2;{\mathbb Z}) \: \cong \: \left( SL(2,{\mathbb Z}) \times
SL(2,{\mathbb Z}) \right) / {\mathbb Z}_2,
\end{displaymath}
where one $SL(2,{\mathbb Z})$ acts on complex structures and the other
on K\"ahler structures.  As anticipated above, the largest part of 
the T-duality group is therefore two $SL(2,{\mathbb Z})$ factors of
Dehn twists, for complex and K\"ahler structures separately.

Let us consider the $SL(2,{\mathbb Z})$ action on the complex structure
of the target-space $T^2$.
The worldsheet fermions transform
in the pullback of the tangent bundle on the target space,
hence (up to dualizations and complex conjugations), they transform
as $dz$, not $\sqrt{dz}$, just as the worldsheet bosons (consistent with
worldsheet supersymmetry).  As a result, it is $SL(2,{\mathbb Z})$,
not $Mp(2,{\mathbb Z})$, that acts on worldsheet fermions themselves 
under T-duality on the target space.

The Ramond sector
vacua, on the other hand, are
a different matter.  The Ramond sector vacua of a physical theory transform
in principle as sections of the pullback of the canonical bundle on
the target space, {\it i.e.} as $\sqrt{dz}$.  (When the target is
Calabi-Yau, this can be subtle to see, but is much more manifest in
other cases, see for example \cite{rflat} for a recent discussion of Fock
vacua transforming as sections of bundles.)
Thus, in principle, the Ramond sector vacua should pick up factors of
$\pm \sqrt{c \tau + d}^{-1}$ under the action of T-duality on the target space,
and in particular transform under $Mp(2,{\mathbb Z})$ rather than
$SL(2,{\mathbb Z})$.  (Similarly, though less relevantly for our purposes,
target-space spin structures are also encoded in the R sector vacua,
in signs under target-space periodicities.)
Since the ${\mathbb Z}_2$ extension acts only on the vacua, it should
commute with the GSO projection, and so our analysis should have no effect
on physical states.

In passing, we should note that the transformation of the Ramond
sector vacua under target-space Dehn twists is the worldsheet realization
of the target-space mapping class group action on fermions discussed
in section~\ref{sect:mappingclass-tori}.

T-duality for higher-dimensional tori can be treated in a very similar
fashion.  Under $SL(n,{\mathbb Z})$ factors of Dehn twists, the worldsheet
fermions themselves will be well-defined, but the Ramond sector vacua
will pick up ambiguous signs (in addition to permutations of the spin
structures).  For the same reasons as in our discussions of mapping class
groups of tori, the Ramond sector vacua are well-defined under the
action of the ${\mathbb Z}_2$ extension $\widetilde{SL}(n,{\mathbb Z})$.

Now, let us ask a different question:  how do the worldsheet fermions
transform under Dehn twists of the worldsheet itself?
Of course, Dehn twists will permute the worldsheet spin structures,
but let us consider possible additional phase factors picked up the
fields themselves.
In principle, the worldsheet fermions in a physical (untwisted) theory
with worldsheet $\Sigma$ are sections of $\sqrt{K_{\Sigma}}$,
hence transform as $\sqrt{dz}$ for $z$ a coordinate on the worldsheet.
As a result, it is natural to propose that
the worldsheet fermions transform under $Mp(2,{\mathbb Z})$
under the action of worldsheet Dehn twists.

\subsection{Moduli spaces of SCFTs}

Let us consider first moduli spaces of SCFTs constructed as sigma models on
elliptic curves.  As we just argued in the preceding subsection, 
although the worldsheet fermions themselves are single-valued under the
action of $SL(2,{\mathbb Z})$ on the target space, the R sector vacua are
not.  In addition to the expected permutations of spin structures, the
R sector vacua also necessarily pick up phases, which due to a square root
branch cut are ambiguous under $SL(2,{\mathbb Z})$.  As a result,
it is natural to propose that
the physically-relevant moduli space of complex structures, half of the
moduli space of SCFTs for elliptic curves, should be of the form
\begin{displaymath}
\left[ \left( \mbox{upper half plane}\right) / Mp(2,{\mathbb Z}) \right],
\end{displaymath}
and so the full moduli space, taking into account both complex and
K\"ahler structures, is of the form
\begin{displaymath}
\left[ \left( \mbox{upper half plane}\right) / Mp(2,{\mathbb Z}) \right]
\times
\left[ \left( \mbox{upper half plane}\right) / Mp(2,{\mathbb Z}) \right].
\end{displaymath}
(We have only discussed the role of the metaplectic group on the space
of complex structures, but by mirror symmetry, analogous considerations must
also apply to the K\"ahler structures.)

This sheds new light on a result in \cite{mebw}.  There, it was argued that
the Bagger-Witten line bundle over a moduli space of SCFTs for elliptic curves
was only well-defined over the stack above, an $Mp(2,{\mathbb Z})$ quotient
of the upper half plane rather than $SL(2,{\mathbb Z})$ or 
$PSL(2,{\mathbb Z})$.  
Now, the Bagger-Witten line bundle over any moduli space of SCFTs is a 
(possibly fractional) line bundle of spectral flow operators, that play
an essential role in target-space supersymmetry.  Here, we have proposed that
the $Mp(2,{\mathbb Z})$ quotient (as opposed to an $SL(2,{\mathbb Z})$ or
$PSL(2,{\mathbb Z})$ quotient) is necessary in order to make the R sector
vacua well-defined.

Analogous results hold in higher dimensions.  For simplicity, let us consider
the space of complex structures on a torus $X$ that is a complex projective
manifold, {\it i.e.} an abelian variety, and has complex dimension $g$,
say.
For a fixed choice of polarization (K\"ahler form) on $X$, 
the complex structures preserving
the polarization are parametrized by a quotient
\begin{displaymath}
A_g  \: = \: H_g / Sp(2g,{\mathbb Z}),
\end{displaymath}
where $H_g$ is the higher-dimensional Siegel upper half plane (meaning,
symmetric $g \times g$ complex matrices with positive-definite imaginary
part), and $Sp(2g,{\mathbb Z})$ is the symplectic group of integral
$2g \times 2g$ matrices preserving the polarization -- the ordinary
symplectic group $Sp(2g,{\mathbb Z})$ for nondegenerate cases
(more properly, principal polarizations).  (See {\it e.g.}
\cite{gh}[section 2.6] for more information.)
Here also, for much the same
reasons, the spin structures are parametrized by the stacky quotient
\begin{displaymath}
\left[ H_g / Mp(2g,{\mathbb Z}) \right],
\end{displaymath}
where $Mp(2g,{\mathbb Z})$ is the metaplectic group extending
$Sp(2g,{\mathbb Z})$ by ${\mathbb Z}_2$.

In general, for an $n$-dimensional torus $T^n$, the group of all
T-dualities is $O(n,n;{\mathbb Z})$; see {\it e.g.}
\cite{gmr} for a thorough discussion, hence ordinarily the complete
moduli space of tori is described as a quotient by $O(n,n;{\mathbb Z})$.  
We will not try to give a thorough description of the precise metaplectic
replacement here, but we will note that
elements of $O(n,n;{\mathbb Z})$
whose determinant is different from one also modify the worldsheet GSO
projection (see {\it e.g.} \cite{gmr,dhs,dlp}, \cite{opu}[section 3.2]),
for example exchanging type IIA and IIB, 
so to give a thorough description of the moduli space of SCFTs with
target an $n$-dimensional torus $T^n$ will presumably involve a
${\mathbb Z}_2$ extension of $SO(n,n;{\mathbb Z})$ (as well as a means of
following spin structures).

\section{Ten-dimensional IIB S-duality}
\label{sect:10d-sduality}

In this section, we will describe the action of S-duality on fermions
in ten-dimensional IIB strings, and we will propose that $SL(2,{\mathbb Z})$
should in principle be replaced by the metaplectic group $Mp(2,{\mathbb Z})$.
(The result of this section has been independently obtained by
D.~Morrison \cite{davepriv}.)

Classically, recall \cite{schwarz-west,howe-west,schwarz} 
that type IIB supergravity in ten dimensions
has an $SU(1,1;{\mathbb C}) \cong SL(2,{\mathbb R})$
symmetry (modulo finite factors) extending a local $U(1)$ symmetry.
In that theory, the fermions 
only transform under the local $U(1)$ (see {\it e.g.}
\cite{schwarz-west}[equ'n (7c)]).
After gauge-fixing,
the local transformation is determined so as to preserve the gauge,
and so an $SL(2,{\mathbb R})$ transformation determines an action
on the fermions \cite{gabgreen}.
(See also \cite{westrev}[section 13.5] for a review.)
Furthermore, in the quantum theory, the continuous
$SL(2,{\mathbb R})$ symmetry is replaced by a local $SL(2,{\mathbb Z})$,
and it is in this fashion that we can understand that the fermions transform
under the action of S-duality.

In the conventions of \cite{gabgreen}, if we pick gauge-fixing
condition
$\hat{\phi}=0$, then under the $U(1)$ subgroup of $SL(2,{\mathbb R})$,
the (complex)
gravitino $\psi_{\mu}$ transforms as \cite{gabgreen}[equ'n (2.11)]
\begin{displaymath}
\psi_{\mu} \: \mapsto \: \exp(i \Sigma/2) \psi_{\mu},
\end{displaymath}
and the (complex) dilatino $\lambda$
transforms as \cite{gabgreen}[equ'n (2.11)]
\begin{displaymath}
\lambda \: \mapsto \: \exp(- 3 i \Sigma / 2) \lambda,
\end{displaymath}
where
for
\begin{displaymath}
\left[ \begin{array}{cc}
a & b \\
c & d \end{array} \right] \: \in \: SL(2,{\mathbb R}),
\end{displaymath}
the $U(1)$ rotation is defined by
\begin{displaymath}
\exp(i \Sigma) \: = \: \left( \frac{ c \overline{\tau} + d }{
c \tau + d} \right)^{1/2}.
\end{displaymath}
Put more simply, and restricting to the local discrete gauge symmetry
$SL(2,{\mathbb Z})$,
\begin{eqnarray*}
\psi_{\mu} & \mapsto &
\left( \frac{ c \overline{\tau} + d }{
c \tau + d} \right)^{1/4} \psi_{\mu}, \\
\lambda & \mapsto & 
\left( \frac{ c \overline{\tau} + d }{
c \tau + d} \right)^{-3/4} \lambda .
\end{eqnarray*}
These transformations were used in {\it e.g.} \cite{greensethi} to argue
that the coefficients of certain higher-dimension operators should behave
as nonholomorphic analogues of modular forms.

Now, as written, the transformations above appear to be invariant under
the center of $SL(2,{\mathbb Z})$.  However, there is a potential subtlety
present in the one-quarter-root branch cuts.  
If we perform
a field redefinition, 
we can construct new fermions
with cleaner transformation laws,
which is what we will describe next.

Following \cite{borcherds}, one can define a modular form of
weight $(m^+,m_-)$ to be a real analytic function $F$ on the upper half
plane $H$ such that
\begin{displaymath}
F\left( \frac{ a \tau + b}{c \tau + d} \right) \: = \:
(c \tau + d)^{m^+} (c \overline{\tau} + d)^{m_-} F(\tau).
\end{displaymath}
For example, ${\rm Im}\, \tau$ is such a modular form, of weights
$(-1,-1)$.
Given any such $F$, one can transform it to a modular form $F'$ of weight
$(m^+-m^-,0)$ defined by
\begin{displaymath}
F'(\tau) \: \equiv \: ({\rm Im}\, \tau)^{m^-} F.
\end{displaymath}
In the present case, $\psi_{\mu}$ transforms like a modular
form of weights $(-1/4,+1/4)$ and $\lambda$ a modular form of weights
$(+3/4,-3/4)$, so we define
\begin{displaymath}
\psi'_{\mu} \: \equiv \: ( {\rm Im}\, \tau )^{1/4} \psi_{\mu},
\: \: \:
\lambda' \: \equiv \: ( {\rm Im}\, \tau)^{-3/4} \lambda.
\end{displaymath}
Then, under the action of $SL(2,{\mathbb Z})$, these new fields transform as
\begin{eqnarray*}
\psi'_{\mu} & \mapsto &
\pm ( c \tau + d )^{-1/2} \psi'_{\mu}, \\
\lambda' & \mapsto &
\pm ( c \tau + d )^{+3/2} \lambda' .
\end{eqnarray*}
Furthermore, we are free to redefine the fermions as we wish --
$\psi'_{\mu}$, $\lambda'$ are no more or less physical than
$\psi_{\mu}$, $\lambda$, so our field redefinition has merely made
more manifest a subtle symmetry of the theory.

The transformations above are only defined up to signs -- the usual ambiguity
in square roots.  As a result, the action of $SL(2,{\mathbb Z})$
is ambiguous, as an element of $SL(2,{\mathbb Z})$ does not uniquely
determine a choice of sign.  The group that really is acting is some
two-fold cover of $SL(2,{\mathbb Z})$.  Given the form of the
transformations above, it is natural to propose that the full duality
group is
the metaplectic group $Mp(2,{\mathbb Z})$.

\section{Four-dimensional $N=4$ SYM}
\label{sect:4dN=4}

Four-dimensional $N=4$ super-Yang-Mills can be directly connected to
ten-dimensional IIB string theory by virtue of the AdS/CFT correspondence.
In particular, in \cite{bgs} it was observed that as the components of
the $N=4$ supercurrent multiplet couple to the fields of the
ten-dimensional IIB supergravity, the transformation properties of the
ten-dimensional fields under $SL(2,{\mathbb Z})$ (or, as we have
observed here, $Mp(2,{\mathbb Z})$) imply transformation laws for the
four-dimensional multiplets.

For example, as discussed in section~\ref{sect:10d-sduality},
if the ten-dimensional dilatino $\lambda$ has weights
$(+3/4,-3/4)$, then the `dual' four-dimensional multiplet denoted
$\Lambda^i_{\alpha}$ in \cite{bgs} ($\alpha \in \{1,2\}$,
$i \in \{1, \cdots, 4\}$) transforms similarly under
$SL(2,{\mathbb Z})$.  We argued
in section~\ref{sect:10d-sduality} that a field redefinition of the
ten-dimensional theory makes the sign ambiguity and extension to
an $Mp(2,{\mathbb Z})$ action more manifest; similar remarks should apply here.

Given the role of $Mp(2,{\mathbb Z})$ we have discussed in this section,
one might ask what happens after compactification of the four-dimensional
theory on a curve.  Such compactifications were
discussed historically in \cite{hms,bjsv}, and more recently
in {\it e.g.} \cite{edgl}.  For example, these papers argued that the
$SL(2,{\mathbb Z})$ of the four-dimensional $N=4$ theory becomes T-duality
in the two-dimensional theory.  Briefly, if the fermions of the
four-dimensional theory transform under the ${\mathbb Z}_2$
extension $Mp(2,{\mathbb Z})$ of $SL(2,{\mathbb Z})$, then the same
must be true of the fermions in the two-dimensional theory, which is consistent
with observations in section~\ref{sect:tduality} 
regarding T-duality and fermions.

In passing, for completeness we should also mention that $SL(2,{\mathbb Z})$
actions on fermions in four-dimensional
$N=2$ $U(1)$ gauge theory 
are discussed
in \cite{edsdualabel}.  In the conventions of that reference,
$SL(2,{\mathbb Z})$ acts honestly on the fermions, but the zero modes
and partition function pick up factors which could have square-root
sign ambiguities.
Specifically, if the fermions into modes of
R-charge $R=1$, denoted $\alpha$, corresponding to a pair of
positive-chirality gluinos, and conjugate fields of
$R=-1$, denoted $\overline{\alpha}$ and of negative chirality, then
under $\tau \mapsto - 1/\tau$,
\begin{displaymath}
\alpha \: \mapsto \: \alpha_D \equiv \tau \alpha, \: \: \:
\overline{\alpha} \: \mapsto \: \overline{\alpha}_D \equiv 
\overline{\tau} \overline{\alpha}.
\end{displaymath}
The zero modes are more nearly relevant for our purposes.
The normalized integration measure for any fermi
zero mode $\beta$ is of the form \cite{edsdualabel}[equ'n (3.15)]
\begin{displaymath}
\frac{ d \beta }{ \sqrt{{\rm Im}\, \tau} },
\end{displaymath}
so that under $\tau \mapsto - 1/\tau \equiv \tau_D$,
\cite{edsdualabel}[equ'n (3.16)]
\begin{eqnarray*}
\frac{d \alpha}{\sqrt{ {\rm Im}\,\tau } } & \mapsto &
\frac{ d\alpha_D}{\sqrt{ {\rm Im}\, \tau_D} } \: = \:
\sqrt{ \frac{\overline{\tau}}{\tau} }
\frac{d \alpha}{\sqrt{ {\rm Im}\,\tau } }, \\
\frac{ d \overline{\alpha} }{ \sqrt{ {\rm Im}\, \tau} } & \mapsto &
\frac{ d \overline{\alpha}_D }{\sqrt{ {\rm Im}\, \tau_D} } \: = \:
\sqrt{ \frac{ \tau }{\overline{\tau} } } 
\frac{ d \overline{\alpha} }{ \sqrt{ {\rm Im}\, \tau} }
\end{eqnarray*}
As a result, since the number of $\alpha$ zero modes minus the
number of $\overline{\alpha}$ zero modes is $- (\chi+\sigma)/2$,
the fermion measure picks up a factor of \cite{edsdualabel}[equ'n (3.17)]
\begin{displaymath}
\left( \frac{ \overline{\tau} }{\tau} \right)^{- (\chi+\sigma)/4}
\end{displaymath}
under $\tau \mapsto - 1/\tau$.
However, this is not the only part of the path integral measure that
transforms.  In particular, the gauge measure also picks up a factor of
\cite{edsdualabel}[equ'n (3.18)]
\begin{displaymath}
\tau^{+(\chi-\sigma)/4} \overline{\tau}^{+(\chi + \sigma)/4}
\end{displaymath}
under $\tau \mapsto - 1/\tau$.
(The integration measure of $a$, $a_D$ is invariant.)
As a result, altogether the path integral for the $N=2$ U(1) super
Yang-Mills theory
picks up a factor of \cite{edsdualabel}[equ'n (3.19)]
\begin{displaymath}
\tau^{- \chi/2}
\end{displaymath}
under $\tau \mapsto - 1/\tau$.
We leave a more detailed examination of four-dimensional
$N=2$ theories and the role of the metaplectic group for future work.

\section{U-duality}
\label{sect:uduality}

In this section we will discuss U-dualities appearing in 
toroidally-compactified theories, which will not only allow us to display
how U-duality groups are modified when one takes into account fermions,
but also perform cross-checks of our proposals.
Briefly, in each dimension we will propose a ${\mathbb Z}_2$ extension
of the ordinary U-duality group, reflecting the fact that taking fermions
into account should only generate sign ambiguities and hence 
we expect only an
overall ${\mathbb Z}_2$ extension, rather than an extension by a larger
finite group. 

\subsection{Nine dimensions}

U-duality groups of nine-dimensional theories were discussed in
{\it e.g.} \cite{edduality,aspu}.  Briefly, a nine-dimensional theory
can be obtained as either M theory on $T^2$ or, equivalently, type II on
$S^1$.  As M theory on $T^2$, for the reasons discussed
in section~\ref{sect:mappingclass-tori}, 
this theory has an $Mp(2,{\mathbb Z})$ duality
when one takes into account fermions.
If we think about this as type IIB on
$S^1$, then as discussed in section~\ref{sect:10d-sduality},
the ten-dimensional type IIB theory has an $Mp(2,{\mathbb Z})$
symmetry,
which coincides with the M theory U-duality group.
In addition, there is also the ordinary T-duality on $S^1$, but as this
exchanges IIA and IIB (albeit shifting the dilaton in the process),
it does not contribute to the duality group of IIB
{\it per se}.

In any event, regardless of how we construct the nine-dimensional
theory, we see that when taking into account the fermions, in
our proposal it has
an $Mp(2,{\mathbb Z})$ symmetry, slightly enlarging what was previously
described as an $SL(2,{\mathbb Z})$ symmetry.  We also see that
the results of section~\ref{sect:mappingclass-tori} and
section~\ref{sect:10d-sduality} constrain one another:  consistency
of the nine-dimensional theory requires the two duality groups obtained
independently in those sections to match, as indeed they do.

\subsection{Eight dimensions}  \label{sect:udual:8d}

In this section we shall discuss the eight-dimensional theory which
can be obtained alternatively as
a compactification of M theory on $T^3$, or of type IIA/B on $T^2$.

In the past, omitting fermions, it was said that the U-duality group
is $SL(3,{\mathbb Z}) \times SL(2,{\mathbb Z})$.  The $SL(3,{\mathbb Z})$
factor arose from the mapping class group of the $T^3$ in the
M theory compactification, and as explained in {\it e.g.}
\cite{opu}[section 4.3], the other factor arises from T-duality of the
type II compactification on $T^2$.  

Omitting fermions, the (GSO-preserving) T-duality
group acting on the SCFT is
\begin{displaymath} 
SO(2,2;{\mathbb Z}) \: \cong \: \left( SL(2,{\mathbb Z}) \times
SL(2,{\mathbb Z}) \right)/{\mathbb Z}_2.
\end{displaymath}  
However, in the spacetime theory,
we cannot\footnote{We would like to thank E.~Witten for this
observation.} quotient out the ${\mathbb Z}_2$, as it acts nontrivially
on the RR fields.  In fact, to better understand this statement,
let us summarize the actions of various
$SL(2,{\mathbb Z})$'s appearing.  The ten-dimensional $SL(2,{\mathbb Z})_S$
of IIB (omitting fermions) acts as \cite{basu1,kp1,schwarz10d}
\begin{eqnarray*}
\tau & \mapsto & \frac{a \tau + b}{c \tau + d}, \\
\left[ \begin{array}{c} B_N \\ B_R \end{array} \right] & \mapsto &
\left[ \begin{array}{cc}
d & -c \\ -b & a \end{array} \right] 
\left[ \begin{array}{c} B_N \\ B_R \end{array} \right], \\
B_R + \tau B_N & \mapsto & \frac{ B_R + \tau B_N }{c \tau + d},
\end{eqnarray*}
where $\tau$ is the complexified ten-dimensional string coupling, $B_N$ is
the ten-dimensional NS-NS $B$ field, and $B_R$ is the R-R $B$ field.
Part of the T-duality group of the IIB compactification on $T^2$, 
which we shall denote $SL(2,{\mathbb Z})_T$,
acts as ({\it e.g.} \cite{basu1})
\begin{eqnarray*} 
\tilde{T} & \mapsto & \frac{ a \tilde{T} + b}{c \tilde{T} + d}, \\
\rho & \mapsto & \frac{\rho}{c \rho + d},
\end{eqnarray*}
for $\tilde{T} = B_N + i V_2$ the $T^2$ K\"ahler modulus, 
with $B_N$ the NS-NS two-form on $T^2$ and $V_2$ the
volume of the $T^2$, and $\rho = - B_R + i \tau_1 V_2$.
The other half of the T-duality group we denote $SL(2,{\mathbb Z})_U$, and
acts on the complex structure modulus $U$ of $T^2$ as ({\it e.g.} \cite{basu1})
\begin{displaymath}
U \mapsto \frac{a U + b}{c U + d}.
\end{displaymath}

From the transformation law of $\rho$ under $SL(2,{\mathbb Z})_T$, we see
that the RR 2-form $B_R$ picks up a sign under the ${\mathbb Z}_2$
center of $SL(2,{\mathbb Z})_T$, and so in forming the string duality group,
we must lift $SO(2,2;{\mathbb Z})$ to a double cover, namely
$SL(2,{\mathbb Z})_T \times SL(2,{\mathbb Z})_U$.

The $SL(2,{\mathbb Z})_S$, $SL(2,{\mathbb Z})_T$ are combined as a pair of
$2 \times 2$ blocks inside a $3 \times 3$ matrix to form the $SL(3,{\mathbb Z})$
factor in the U-duality group, which altogether is $SL(3,{\mathbb Z}) \times
SL(2,{\mathbb Z})_U$.

We claim that, when fermions are taken into account,
the U-duality group in this theory is 
$( \widetilde{SL}(3,{\mathbb Z}) \times Mp(2,{\mathbb Z}) )/{\mathbb Z}_2$.
The $\widetilde{SL}(3,{\mathbb Z})$ factor arises from the mapping
class group of $T^3$ in the M theory
compactification, as in section~\ref{sect:mappingclass-tori}.

In any event, there would appear to be two natural possibilities for
the U-duality group in eight dimensions:  either
$\widetilde{SL}(3,{\mathbb Z}) \times Mp(2,{\mathbb Z})$, or
$ ( \widetilde{SL}(3,{\mathbb Z}) \times Mp(2,{\mathbb Z}) )/{\mathbb Z}_2$,
where the ${\mathbb Z}_2$ quotient acts on the two central 
${\mathbb Z}_2$ extension factors.  As we are looking for a 
${\mathbb Z}_2$ extension of the ordinary U-duality group rather than a 
${\mathbb Z}_2 \times {\mathbb Z}_2$ extension, we propose that the
correct U-duality group in eight dimensions is
\begin{displaymath}
\frac{  \widetilde{SL}(3,{\mathbb Z})
\times Mp(2,{\mathbb Z})
}{
{\mathbb Z}_2
}.
\end{displaymath}
We shall also see that this group arises in the decompactification limit
from seven dimensions in the next section.

\subsection{Seven dimensions}

In this section, we will discuss the U-duality group of the
theory which can alternatively be described as M theory compactified
on $T^4$, or
as type II on $T^3$.

In the past, omitting fermions, it was said that the U-duality group
is $SL(5,{\mathbb Z})$.  The mapping class group of the $T^4$ appearing
in the M theory compactification, namely $SL(4,{\mathbb Z})$,
appears as a subgroup of $SL(5,{\mathbb Z})$ (embedded in the obvious
way, as a $4 \times 4$ block inside $5 \times 5$ matrices
\cite{edduality}[section 3.4]).

Similarly (see {\it e.g.} \cite{opu}[section 4.3]),
the (GSO-preserving and RR-compatible)
T-duality group $SL(4,{\mathbb Z})$
of the type II compactification
does not commute with the mapping class group of the M theory compactification,
and the two of them together generate $SL(5,{\mathbb Z})$.

In the present case, taking into account fermions, our proposal is that
the contribution to
the U-duality group from the mapping class group
arising from M theory on $T^4$ is the ${\mathbb Z}_2$ extension
$\widetilde{SL}(4,{\mathbb Z})$ of $SL(4,{\mathbb Z})$, as discussed
in section~\ref{sect:mappingclass-tori}.
Similarly, we expect (though have not carefully checked) that the
relevant T-duality group is a 
(different) $\widetilde{SL}(4,{\mathbb Z})$.
They should combine into a ${\mathbb Z}_2$ extension of
$SL(5,{\mathbb Z})$, and the natural possibility is 
$\widetilde{SL}(5,{\mathbb Z})$, which we conjecture to be the case.

Formally repeating the arguments in \cite{edduality}[section 3.4],
the U-duality group of the eight-dimensional decompactification limit
should be the subgroup
\begin{displaymath}
\frac{  \widetilde{SL}(3,{\mathbb Z})
\times Mp(2,{\mathbb Z})
}{
{\mathbb Z}_2
},
\end{displaymath}
a ${\mathbb Z}_2$ extension of $SL(2,{\mathbb Z}) \times
SL(3,{\mathbb Z})$, consistent with our results in section~\ref{sect:udual:8d}.

\section{Conclusions}

In this short paper we have proposed that duality group actions in
high-dimensional theories with maximal supersymmetry should be
slightly enlarged, by nontrivial ${\mathbb Z}_2$ extensions, 
to correctly describe
duality group actions on fermions.  We have argued this separately
for mapping class groups of tori in M theory compactifications,
T-duality groups, ten-dimensional IIB S-duality, and briefly
four-dimensional $N=4$ theories, and checked the proposals against one
another by exploring U-duality groups in dimensions nine, eight,
and seven.

We have only considered U-duality groups of
high-dimensional string compactifications.  It would be interesting,
albeit more technically complex, to extend to lower-dimensional cases
and cases with less supersymmetry.

One possible application of such an extension would be to try to
identify the Bagger-Witten bundle over moduli spaces of K3 superconformal
field theories.  One of the original motivations for this work,
after all, was to understand whether the metaplectic group appearing
in the description of the stringy moduli stack
of elliptic curves given in \cite{mebw} was merely a formal quirk or
reflected physical dualities.
An understanding of U-duality groups in six-dimensional compactifications
of string theory could be used to give analogous information concerning
moduli stacks of K3 superconformal field theories.

Another natural question concerns spinors and Bagger-Witten line bundles
on special K\"ahler moduli spaces.
As briefly outlined in section~\ref{sect:review}, the analogue of a spin
structure on a symplectic manifold is a metaplectic structure, defining
a bundle whose structure group is a metaplectic group.  This suggests
that a detailed investigation of spinors and Bagger-Witten line bundles
on special K\"ahler manifolds
will reveal that the metaplectic group plays an important role there
as well.  We leave such considerations for future work.

In passing, it is also tempting to wonder whether considerations such as those
in this paper for $N=4$ theories in four dimensions are relevant to
metaplectic geometric Langlands theory, as in \cite{edgl,gl1}.

\section{Acknowledgements}

We would like to thank L.~Anderson, P.~Aspinwall,
D.~Auroux, J.~Distler, J.~Francis,
J.~Gray, I.~Melnikov,
D.~Morrison, B.~Pioline, S.~Sethi, P.~West, and E.~Witten for useful
conversations.  
This work was performed in part at the Aspen Center for Physics, which
is supported by National Science Foundation grant PHY-1066293.
T.P. was partially supported by NSF grants DMS-1302242 and
DMS-1601438.
E.S. was partially supported by NSF grant PHY-1417410.

\end{document}